\documentclass[12pt]{iopart}

\usepackage{amssymb}
\usepackage{epsfig}
\usepackage{setstack}
\newcommand{\be}{\begin{equation}}
\newcommand{\ee}{\end{equation}}
\newcommand{\bea}{\begin{eqnarray}}
\newcommand{\eea}{\end{eqnarray}}

\begin{document}

\title{Cosmology Without Averaging}

\author{Timothy Clifton\\
\qquad \vspace{-7pt}}

\address{Department of Astrophysics, Keble Road, University of Oxford,
  UK\\
\qquad \vspace{-7pt}\\
and\\
\qquad \vspace{-7pt}\\
Theoretical Physics Group, CERN, Geneva, Switzerland.}

\eads{\mailto{\mailto{t.clifton@cantab.net}}}


\begin{abstract}

We construct cosmological models consisting of large numbers of
identical, regularly spaced masses.  These models do not rely
on any averaging procedures, or on the existence of a global Friedmann-Robertson-Walker
(FRW) background.  They are solutions of Einstein's equations
up to higher order corrections in a perturbative expansion, and have
large-scale dynamics that are well modelled by the Friedmann equation.
We find that the existence of arbitrarily large density contrasts does not
change either the magnitude or scale of the background expansion,
at least when masses are regularly arranged, and up to the prescribed level of accuracy.
We also find that while the local space-time geometry inside each
cell can be described as linearly perturbed FRW, one could argue that
a more natural description is that of perturbed Minkowski space (in which case the
scalar perturbations are simply Newtonian potentials).
We expect these models to be of use for understanding and testing ideas about
averaging in cosmology, as well as clarifying the relationship between
global cosmological dynamics and the static space-times associated
with isolated masses.

\end{abstract}

\maketitle

\section{Introduction}

The Friedmann-Robertson-Walker (FRW) model is ubiquitous in modern
cosmology, and is widely believed to be a good model for a universe
with a matter content that is approximately uniformly distributed
over large scales.  FRW has had great success in
fitting a variety of cosmological observations, but is not without its problems.
In particular, it has the apparent defect of requiring large
amounts of dark matter and dark energy in order to be compatible with observations.
More fundamentally, it relies on the implicit
assumption that it is permissible to use non-local `average' energy
densities in Einstein's equations, which are a set of local field equations.
The difficulties involved with this are exacerbated by the diffeomorphism covariance of the theory, which
means that there is in general no preferred set of space-like
hypersurfaces with which to perform an average over at all.  What is more, 
even if a suitable and unique averaging scheme is found, it seems we
will still be left with a back-reaction effect due to the non-commutativity of averaging and
evolution under Einstein's equations \cite{ellis,buchert}.  These issues
require further study in order to be fully understood.

One way to make progress in this area is to construct alternative
models that are approximately homogeneous
and isotropic on the largest scales, but that do not involve any
averaging procedures.  We will survey some of the relevant literature
on progress toward this goal in the section that follows, and then
proceed to present our own approach to constructing a cosmological model that is
composed of discrete masses, rather than a continuous fluid.  This
model will appear homogeneous and isotropic when coarse-grained over the largest
scales, but will also be able to accommodate arbitrarily large local density contrasts.  The
validity of this model will be based solely on Newtonian and
post-Newtonian perturbative expansions about Minkowski space,
which are generally considered to be able to easily model large density
gradients without breaking down (for a recent discussion of
the potential difficulties involved in modelling the corresponding situation
with a perfect fluid filled, perturbed FRW cosmology see, e.g., \cite{ras}).

The model we construct is one in which isolated masses are arranged in
a regular array, which by application of the Israel junction
conditions is a solution of Einstein's equations up to a specified
level of accuracy.  Contrary to some previous results in the
literature, we show that discretisation of the matter content does {\it not} affect the
global expansion of the space-time, which proceeds in just the same
way as a perfect fluid FRW solution with the same energy density and
spatial curvature, up to the required accuracy.  This result says
nothing about what happens in the more realistic situation of
irregularly arranged masses, but does show that, in at least some
situations, perturbed FRW is a good description of a universe with
(arbitrarily) large density contrasts.  Establishing the existence of
such situations is of interest not only for limiting the possible consequences of
inhomogeneity on the large-scale expansion of the Universe, but also
for testing the viability of proposed methods of accounting for
more general inhomogeneity in the Universe:  If they predict modifications to the
large-scale expansion in configurations where it is known that none
should occur, then one may choose to question their viability.

In Section \ref{2b} we survey some of the literature on inhomogeneous
cosmological models, concentrating in particular on studies that have
similarities to the approach used here, and highlighting predictions
of deviations from the usual Friedmann expansion.  In Section \ref{3}
we present the type of model we will be considering, and the perturbative expansion we
deploy is outlined in Section \ref{pertsec}.  Section \ref{statsec}
treats the geometry inside each of the primitive cells of our structure as
being perturbed Minkowski space, and finds the corresponding
cosmological evolution, after the Israel junction conditions have been
applied.  A similar analysis is then performed in Section
\ref{timesec}, this time around a time dependent background.  In
Section \ref{7} we relate the time independent and time dependent
approaches used in the preceding two sections, and in Section \ref{8}
we show that the model we have constructed has identical large-scale
expansion to a perfect fluid filled FRW cosmology with the same energy
density and large-scale spatial curvature.  In Section \ref{obs} we consider the
problem of determining cosmological observables, such as redshifts and
luminosity distances, and in Section \ref{disc} we conclude.

\section{Previous Results}
\label{2b}

Before proceeding with our study, let us briefly review some of the
relevant literature on cosmological models containing discrete masses,
and the affect of structure on the evolution of the Universe.  These
studies have frequently suggested that the presence of inhomogeneity
in the Universe could affect its global expansion rate.

One of the first studies to include discrete masses in a cosmological
model was that of McVittie \cite{mcv}.  Here a space-time geometry similar to the
Einstein static universe was considered, and McVittie concluded that
if instead of having a perfectly homogeneous and isotropic matter
content, it was instead the case that a number, $n$, of singularities
of mass $m$ were allowed to develop, then the corresponding spatial volume of
the universe would be
\be
V=\frac{16 M^3}{\pi} \left( 1 +\frac{2 n m}{M} \right) > \frac{16
    M^3}{\pi} = V_{Einstein},
\ee
where $M$ is the total mass in the universe, and $V_{Einstein}$ is the
spatial volume of the perfect Einstein static universe.  McVittie then
reasoned that because the spatial volume of such a universe was larger
after structures were allowed to form, that this signalled
instability of the Einstein static universe, which requires a
particular value of $V$ in order to exist.  For the present purposes
one could interpret this result as saying that the presence of discrete
structures changes the scale of the cosmological solution.

The next significant development we are aware of, in terms of
cosmological models containing discrete structures, is due to Einstein and
Strauss \cite{es}.  Here the authors embed discrete masses into an FRW
background by excising spherical regions of the homogeneous fluid, and
replacing them with vacuoles containing singularities at their centres.  The space-time inside each
`vacuole' is then well modelled by either Schwarzschild geometry, or
perturbed FRW (provided the vacuole is small compared to the Hubble
scale of the background solution).  The resulting structure is
often referred to as a `Swiss cheese universe', and is an exact solution of
Einstein's equations when Schwarzschild geometry is used.  In these models the 
presence of the singularities does not affect the expansion of the
background FRW space-time.  One should, however, be aware that this is true
by construction:  The vacuoles all have exactly FRW boundary
conditions, and are not allowed to intersect in the usual application
of the model.  Furthermore, one could question whether the requirement
of spherical symmetry restricts the possible behaviours that might
otherwise be possible.  In the present study we will lift the
requirements of perfect FRW boundary conditions and spherical
symmetry, although the models we consider are still highly
symmetric.

\begin{figure}[tbp]
\centerline{\epsfig{file=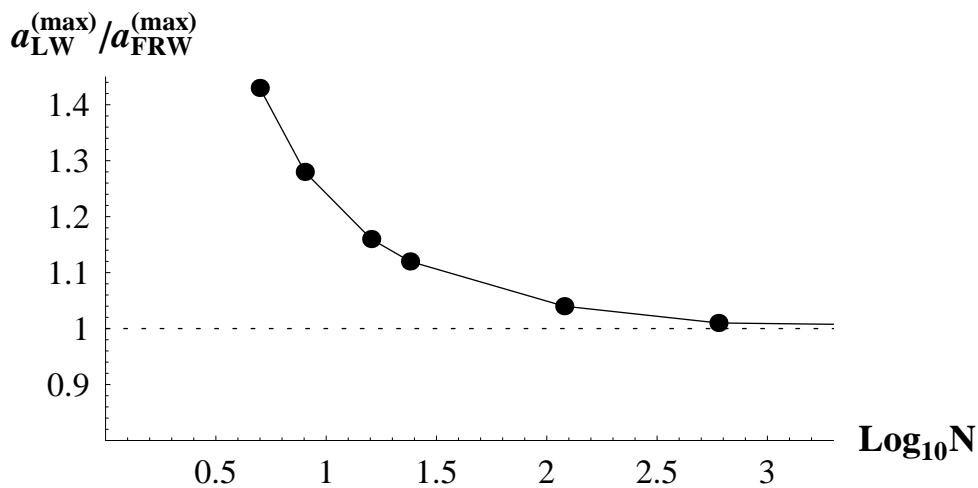,height=6.5cm}}
\caption{\textit{The maximum of expansion in the Lindquist-Wheeler
    model, $a^{(max)}_{LW}$, as a fraction of the corresponding
    maximum in a closed FRW universe with the same energy
    density and spatial curvature, $a^{(max)}_{FRW}$.  The abscissa gives the number of cells in the
    Lindquist-Wheeler lattice, $N$.
}}
\label{lwfig}
\end{figure}
In 1957 the problem of cosmological models containing discrete masses
was returned to by Richard Lindquist and John Archibald Wheeler
\cite{LW}.  These authors attempted to build a cosmological
model in analogy to the Wigner-Seitz construction of
solid-state physics.  The basic idea here is to construct a regular
lattice of cells, and then solve the field equations by approximating
the influence of all cells external to the one under consideration as
being spherically symmetric.  In the context of general relativity
this results in the space-time inside each cell being uniquely
given by Schwarzschild geometry, and unlike the case of solid-state
physics, results in a non-zero normal derivative of the relevant field at
the boundary of each cell.  The lattice therefore undergoes a global
expansion, that can be compared to the Friedmann solutions of FRW
cosmology.  By considering spatially closed lattices Lindquist and
Wheeler deduced that the expansion of their model had the same
functional form as the usual Friedmann solution, but with a different scale of
expansion, as shown in Fig. \ref{lwfig}.  Luminosity distances
and redshifts in this model were calculated in \cite{cf1} and 
\cite{cf2}, where deviations from the corresponding quantities in FRW
cosmology were identified.  The problem of finding an exact $2$-body
solution within this framework has also been addressed recently in \cite{larena}.
While compelling, however, the Lindquist-Wheeler model suffers from the problem of relying on an
approximation scheme that is difficult to quantify.  In the present
study we attempt to produce a similar model to that created by
Lindquist and Wheeler, but with a more clearly defined set of approximations.

More recent studies on the affect of structure on the expansion of the
Universe often come under the title `back-reaction', or `macroscopic
gravity'.  Let us consider the case of macroscopic gravity first.  The
basic idea behind this approach is that when describing an `average'
cosmological evolution we are not necessarily interested in the local
expansion rate any any particular point in the Universe, but rather in
the non-local expansion inferred from observations made over large
distances.  Now, while Einstein's equations are thought to be the
appropriate ones to describe the local curvature of space-time, if we consider
non-local quantities, such as `average' expansion, then we may need a
different set of equations.  The problem of how averaging should best be
performed in general relativity is a difficult, and still somewhat
open question.  Progress toward solving this problem has, however, been made
by Zalaletdinov in his macroscopic gravity theory, which modifies
Einstein's equations by including gravitational correlation correction
terms \cite{zal}.  Applying this approach to cosmological solutions
Coley, Pelavas and Zalaletdinov have found that there exist
homogeneous and isotropic exact solutions to the macroscopic field
equations in which the correlation tensor takes the form of a spatial
curvature term \cite{cpz}.  This result again supports the idea that
the formation of structure in the Universe could lead to a change in
scale of the global expansion.

Finally, let us return to studies of the `back-reaction' problem.  The
basic idea here is that averaging over a space-like hyper-surface, and
evolution of the same hyper-surface under Einstein's equations, do not
commute, so that if we wish to successfully evolve an averaged space
forward in time we should add corrections to Einstein's equations.
The affect of these extra terms is what is known as the back-reaction
of the structures that form in the Universe on the overall expansion,
and is a subject that has recently attracted much attention (see,
e.g., \cite{buch} for a review and references).  It was shown in
\cite{lar1} that for scaling solutions to exist in this approach then
the extra terms in the effective Friedmann equation should either appear as an effective
massless scalar field, or with the same form as the spatial curvature
term.  The former case is inconsequential for the expansion of the late
Universe, but if the latter is true then this once again points to a
change in scale of the global cosmological dynamics (although one
should be aware that the spatial curvature term in the effective
Friedmann equation here does not have to scale as it usually does in
FRW cosmology).

We consider the results of these previous investigations more than sufficient
motivation to further investigate models with a discretized matter
content.  

\section{A Lattice of Discrete Masses}
\label{3}

Our aim is to further develop the ideas outlined above by constructing
a well defined cosmological model in which the Universe is filled
with a large number of identical, regularly spaced masses, but has no
fluid filled background cosmology.  To achieve this
we begin as Lindquist and Wheeler did, by considering 
a number of cells that are regular polytopes, and that
are identical to each another up to spatial translations and
rotations.  At the centre of each cell we then place a
non-rotating, chargeless object of mass $m$.  These cells
will act as the building blocks of our model, and our objective is to
arrange them in such a way as to construct a smooth, geodesically
complete space-time\footnote{Up to the possible existence of
  singularities at the centre of each cell.}.

To satisfy Einstein's equations we must
have that the geometry inside each cell satisfies $R_{a b}=0$ in the
exterior region of the central mass, and that the boundaries between
cells satisfy the Israel junction conditions in vacuum: That the
induced metric and extrinsic curvature of the boundary are the
same on either side \cite{israel,israel2}.  The high degree of
symmetry in the situation we are considering makes these conditions
considerably simpler than they are in general.  Mirror symmetry of any two cells about
their mutual boundary means that the induced metrics on either side of that boundary
are automatically identical.  The conditions of identical extrinsic
curvature are less trivial, but the symmetry of the situation
is again very useful.

\begin{figure}[tbp]
\centerline{\epsfig{file=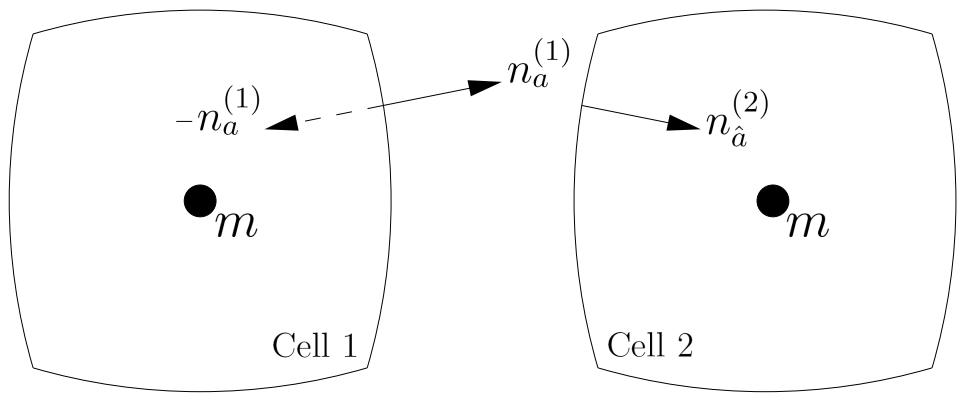,height=5.5cm}}
\caption{\textit{A schematic of the vectors involved in matching two
    cubic cells with central mass $m$.  $n^{(1)}_a$ and $n^{(2)}_{\hat{a}}$ are
    space-like unit vectors normal to the boundary.  The dashed vector
    is $-n^{(1)}_a$, and is mirror symmetric with $n^{(2)}_{\hat{a}}$.}}
\label{n}
\end{figure}

In Figure \ref{n} we show a cross-section of the two cells we are
trying to match in some coordinates $x^{a}$ for the first cell,
and $x^{\hat{a}}$ for the second cell.  The extrinsic curvature in the
first cell is then given by $K^{(1)}_{ab}=n^{(1)}_{a;b}$, where
$n^{(1)}_{a}$ is the space-like unit vector normal to the boundary,
and pointing out of the cell.  In the second cell, the extrinsic
curvature is similarly given as $K^{(2)}_{\hat{a}\hat{b}}=n^{(2)}_{\hat{a};\hat{b}}$, where
$n^{(2)}_{\hat{a}}$ is again the space-like unit vector normal
to the boundary, but this time pointing into the cell.  The covariant
derivatives in these expressions should be taken with respect to the
space-time geometry of the cell in question.  The conditions for
identical extrinsic curvature of the boundary on either side are then
\be
\label{K}
\frac{\partial x^a}{\partial \xi^{i}} \frac{\partial
  x^b}{\partial \xi^{j}} K^{(1)}_{ab} = \frac{\partial x^{\hat{a}}}{\partial \xi^{i}} \frac{\partial
  x^{\hat{b}}}{\partial \xi^{j}} K^{(2)}_{\hat{a}\hat{b}} \equiv K_{ij},
\ee
where $\xi^{i}$ are intrinsic coordinates on the boundary.  From
the symmetry of the situation, however, it can be seen that we could
just as easily have calculated the extrinsic curvature of the boundary
in the second cell by taking the covariant derivative of an inward
pointing normal vector in the first cell, as illustrated in
Figure \ref{n}.  As this vector is given by $-n^{(1)}_a$ 
we must have $K_{ij}=-K_{ij}$
in order to satisfy (\ref{K}), or equivalently
\be
\label{K0}
K_{ij} =0.
\ee
The junction conditions are therefore satisfied if, and only if, the
2+1 dimensional boundary is extrinsically flat.

Now that boundary conditions for each of the cells are known, the
field equations within each cell can be solved.  These are simply
the vacuum Einstein equations, $R_{ab}=0$.  Note, however, that
without boundary conditions these equations do not have a unique
solution.  We therefore need Eq. (\ref{K0}) to find the space-time
geometry inside each cell.  As we will discuss in the next section, we
expect this space-time to be well described by the usual Newtonian and
post-Newtonian approach.  This does not, however, mean that the
entire cosmological model can be described in a single Newtonian
frame-work:  The Newtonian descriptions valid within each cell cannot
describe arbitrarily large numbers of cells simultaneously.  This should be
clear from the fact that on scales of the order of a Hubble length we
expect recessional velocities to approach the speed of light.  Rather, the
way in which the Newtonian descriptions that are valid within the
domain of each individual cell should be related to one another
can be deduced from the boundary conditions, Eq. (\ref{K0}), and will
be spelled out in the sections that follow.

Unlike most approaches to building a cosmological mode, the current
one does not require us to write down one line-element that is
valid for the entire observable Universe.  Instead, due to the
periodicity of the structure we are considering, it is sufficient to
consider only a single cell.  Once we know the geometry, extent, and
rate of expansion of any one cell, we then know the space-time
geometry of the entire universe (up to regions were our
approximations break down, as should be expected, for example, in the
early universe, or near the Schwarzschild radius of a compact object).

One could also, conceivably, consider more complicated structures than
the simple polychora described above.  As long as it can be shown that the
space-time geometry is symmetric about the boundaries between cells, then
the junction conditions will still be satisfied by Eq. (\ref{K0}), and
the global space-time geometry can again be deduced in the manner just
described.  In such a case, however, there may be more than a single type
of primitive cell to consider.  We will not try and construct
such situations here, preferring instead to concentrate on the
simplest structures possible:  Those built from a single repeated polytope.

\section{Perturbative Expansion}
\label{pertsec}

We will not attempt to proceed by looking for an exact solution to
Einstein's equations, as we expect this would be prohibitively difficult.
Instead we will treat the problem perturbatively,
and within each cell will expand with respect to some small parameter:
\be
\epsilon \sim \frac{v}{c}, 
\ee
where $v$ is the typical velocity associated with the type of objects
we will be considering, and $c$ is the speed of light.  For planetary
and galactic systems it is almost always the case that $v/c \lesssim
0.01$.  Furthermore the Newtonian potentials associated with such
systems are nowhere greater than $\phi \sim 10^{-4}$, except within
the vicinity of black holes and neutron stars.  We can therefore assign
\be
\phi \sim \epsilon^2.
\ee
Given that the evolution of these systems are governed by the
motion of their constituents we also have that $\partial/\partial t \sim v
\cdot \nabla$, which implies that
\be
\frac{\vert \partial/\partial t \vert}{\vert \partial/\partial x
  \vert} \sim \epsilon,
\ee
so that time derivatives add an extra order of smallness.  Booking in
orders of smallness in this way is familiar from the usual approach to
post-Newtonian gravitational physics \cite{tegp}. Here, however, we will be
concerned with the vacuum region outside of the central mass of each
of our cells.  We therefore need not consider the orders of smallness
associated with rest mass density, pressure, or any other form of
energy density.

We expect the expansion we have just described to be applicable as
long as the size of each cell is large compared with the Schwarzschild radius
of its central mass, and as long as the number of cells within one
cosmological horizon is also large.  That is, we will be considering
situations in which the bulk of the interior of each cell is well
described by the usual Newtonian and post-Newtonian gravitational
physics.  We will not be concerning ourselves here with what happens near
the singularities that may exist at the centre of each cell, and will
not allow the cells to be so large that their boundaries would appear to have a
recessional velocity any greater than $\sim 0.01 c$.  In fact, we will
have in mind throughout this article cells that are about $1$Mpc wide, with a Milky
Way sized mass at their centre, so that $\phi \sim 10^{-7}$ at the
edge of each cell.  This is well within the limits just mentioned.
For further details of the perturbative expansion used in post-Newtonian physics we
refer the reader to \cite{tegp}.

Let us now consider specifically the motion of our cell boundaries,
whose trajectory we will take to be tangent to the 4-vector $X^a$.  If $n_a$
is normal to this boundary then we can write that the boundary has
4-velocity
\be
u^a \equiv \frac{dX^a}{d\tau} = \frac{dt}{d\tau} \left(1;
\frac{dX^{\mu}}{dt} \right),
\ee
where $\tau$ is proper time along $X^a$, and $\mu$ runs over spatial
indices, and that the condition $u^a n_a=0$ then gives (for $dt/d\tau \neq 0$)
\be
n_t=-n_{\mu} \frac{dX^{\mu}}{dt}.
\ee
Hence, if $dX^{\mu}/dt \sim O(\epsilon )$, so that it has the order of
magnitude associated with a velocity in the perturbative expansion
just outlined, then
\be
\frac{n_t}{n_{\mu}} \sim \epsilon.
\ee
It then follows that $n_t$ has an $O(\epsilon )$ of smallness compared to
$n_{\mu}$, which is expected to be $\sim 1$.  Rather than explicitly
quoting $\epsilon$ in what follows, we will instead simply state that
quantities have a certain order of smallness associated with them.
That this smallness is prescribed by a factor of $\epsilon$
should be taken to be implicit.

\section{Fluctuations About a Static Background}
\label{statsec}

\subsection{Large-scale Evolution Equations}

We can now address the question of the space-time geometry inside
of each cell, and the motion of the cell boundaries that results
from Eq. (\ref{K0}). 
Our initial ansatz for the geometry inside a cell will be linear
perturbations around a Minkowski background.  This is the
standard way to model the weak gravitational fields around massive
objects.  For this, we will use the `conformal Newtonian' gauge, where
the line-element is written
\be
\label{1}
ds^2=-(1+2 \phi)dt^2+(1-2\psi) (dx^2+dy^2+dz^2).
\ee
Here the functions $\phi$ and $\psi$ are gravitational potentials, and
in general relativity we have the well known result that $\psi=\phi$.

Using the perturbative expansion outlined in Section \ref{pertsec}, we
can now write the lowest non-trivial order of each component of the
extrinsic curvature of the boundary at the edge of one of our cells as
\bea
\label{Kmin1}
K_{a b} dx^a dx^b &=& \left( n_{t,t}-n_{\mu} \phi_{,\mu} \right) dt^2 + \left(
n_{\mu,t}+n_{t,\mu} \right) dx^{\mu} dt 
\nonumber \\ && 
+ \left( n_{\mu,\nu} +2 \psi_{,\mu} n_{\nu} - \delta_{\mu ,\nu} \psi_{,\sigma} n_{\sigma}
\right) dx^{\mu} dx^{\nu},
\eea
where indices $\mu, \nu, \sigma$ denote spatial components.  The $K_{tt}$ and
$K_{\mu \nu}$ terms should be expected to have $O(4)$ corrections here, and
the $K_{t \mu}$ and $K_{\mu t}$ terms $O(3)$ corrections.
The time component of the unit vector normal to $\Sigma$ has been
assigned an $O(1)$ of smallness in comparison to the space-like
components here, as discussed in Section \ref{pertsec}.

Now, let us apply the coordinate transformation from Eq. (\ref{K}) to the
expression in Eq. (\ref{Kmin1}), in order to find $K_{ij}$.
To explicitly give the coordinates $\xi^i$ required for this let us first pick out a preferred
space-like direction $x$, which is orthogonal to the boundary at
point where it is at its closest to the central mass.  Such a
direction can always be made to correspond to one of the coordinates in
Eq. (\ref{1}) by performing spatial rotations. The remaining two spatial directions
are then uniquely defined up to a rotation, and we will denote these
directions by the indices $A,B,C$ {\it etc.}.  We can now choose
coordinates on the boundary at $x=X(t,x^A)$ to be given by $\xi^{i} = (t,x^A)$.
Finally, let us define two new derivative operators which act along
the boundary in time-like and space-like directions:
\bea
\dot{\;} &\equiv& u^a \partial_{a} = \partial_t + X_{,t} \partial_x
\nonumber \\ \nonumber 
\:_{\vert A} &\equiv& m^a \partial_{a}= \partial_{A} + X_{,A}
\partial_x,
\eea
where $u^a$ and $m^a$ are time-like and space-like vectors in the
boundary in planes of constant $x^A$ and $t$, respectively.  The
extrinsic curvature of the boundary can then be written in the
coordinate basis of $\xi^i$ as
\bea
\hspace{-50pt}
K_{ij} d\xi^i d\xi^j &=& \left( \dot{n}_{t} +\dot{X}\dot{n}_x -n_{\mu} \phi_{,\mu} \right) dt^2 + \left(
\dot{n}_{A} +n_{t \vert A} +X_{\vert A} \dot{n}_x +\dot{X} n_{x\vert A} \right) dx^{A} dt 
\nonumber \\ && 
+ \Bigg[ n_{A \vert B} + X_{\vert A}n_{x\vert B} - \left( \psi_{,C}
  n_{C}- \psi_{,x} n_x \right) \left( \delta_{AB} + X_{\vert A}
  X_{\vert B} \right) \nonumber \\ &&
\qquad -2 \left( \delta_{AB} \psi_{,x} n_x -\psi_{,A} n_B - \psi_{,x}
  X_{\vert A} n_B -n_x \psi_{,A} X_{\vert B} \right) \Bigg]
dx^{A} dx^{B},\label{Kmin1b}
\eea
where space-time dependent quantities, such as derivatives of $\phi$
and $\psi$, should implicitly be taken to be evaluated on the boundary.
This expression can now be simplified by making use of the 
orthogonality of $n^a$ with respect to $u^a$ and $m^a$.  This gives us
\bea
n_t &=& - n_x X_{,t} \\
n_A &=& - n_x X_{,A}, \label{nA}
\eea
which allows us to write Eq. (\ref{Kmin1b}) as
\bea
\hspace{-50pt}
K_{ij} d\xi^i d\xi^j &=& - n_x \Bigg[ \left(\ddot{X} +\phi_{,x} -
  X_{\vert A} \phi_{,A} \right) dt^2
+ \left( (X_{\vert A})\dot{\;} + (\dot{X})_{\vert A} \right) dx^A dt
\nonumber \\ &&
\qquad \;\;\;+ \Big( X_{\vert AB}+ (\psi_{,x}-X_{\vert C} \psi_{\vert
    C}) (\delta_{AB} +X_{\vert A} X_{\vert B}) \Big) dx^A dx^B
\Bigg].
\label{Kmin2}
\eea
The boundary conditions in Eq. (\ref{K0}) can now be
straight-forwardly applied to (\ref{Kmin2}).  

We find that our lattice is a solution of Einstein's equations (to lowest order in our
perturbative expansion) if the boundary satisfies the conditions:
\be
\label{static}
\ddot{X} = -\sqrt{1+(X_{\vert A})^2} \; (n \cdot \nabla \phi) \vert_X
+O(4)
\ee
and
\be
X_{\vert AB} = (\delta_{AB} +X_{\vert A} X_{\vert B}) \ddot{X}+O(4)
\label{static2}
\ee
together with $(X_{\vert A})\dot{\;} = \big( \dot{X} \big)_{\vert A} =0+O(3)$, where
we have now taken $\psi=\phi$, and where we have used $n^a n_a=1$.  On
the RHS of these equations we have also used the
notation $\nabla = \partial_{\mu}$, and made explicit that the
gradient of $\phi$ should be evaluated at $x=X$.

The potential $\phi$ must of course also satisfy the field equation
$R_{ab}=0$ in the bulk of the cell, which gives us the equations:
\be
\label{statica}
\nabla^2 \phi = 0+O(4)
\ee
and
\be
(\nabla \phi)_{,t} = 0+O(5).
\label{staticb}
\ee
If Eqs. (\ref{static})-(\ref{staticb}) are satisfied, then our
lattice is a cosmological solution to Einstein's equations, up to the
specified order, with arbitrarily large density contrast, and without
any averaging having been performed.

\subsection{Cosmological Solutions}

Above we have derived the evolution equations for a lattice
constructed from a large number of individual cells, each of which
contains an isolated central mass.  
%
In this case the metric fluctuations $\phi$ can be seen to be given
simply by the usual solutions to the Newtonian Poisson equation in
vacuum, with Neumann boundary conditions specified on $X$ via
Eq. (\ref{static}). One can then see from Eq.
(\ref{static}) that if $(n \cdot \nabla \phi) \vert_X >0$ (so that the force due to the
central mass is attractive) then the boundaries must follow
trajectories along which $\ddot{X}<0$.  The global expansion must therefore
always be decelerating.  Eq. (\ref{static2}) then tells us that we must
also have $X_{\vert yy}$ and $X_{\vert zz}<0$, so that in the coordinates
of Eq. (\ref{1}) the boundaries are concave, when viewed from inside the cell
(as is supposed to be implied in Figure \ref{n}).

The first thing that one can now find is the equation of motion of the
element of the boundary that is closest to the central mass of the
cell.  We will label the position of this closest point as $X_c(t)=X(t,0,0)$.
From our choice of coordinate we then have $X_{c\vert
  y}=X_{c\vert z}=0$ and $(n \cdot \nabla \phi)\vert_{X_c}=
\phi_{,x}\vert_{X_c}$, so that Eq. (\ref{static}) becomes
\be
\label{Xc}
\ddot{X}_c = - \phi_{,x}\vert_{X_c} +O(4).
\ee
This boundary element is therefore in free fall in the potential $\phi$, and
the shape of the boundary at this point is given simply by $X_{c\vert y
  y}=X_{c\vert zz} = - \phi_{,x}\vert_{X_c}$.  Given Eq. (\ref{Xc})
for $X_c$, we can now see that the solution for $X$ at all other $y$ and $z$
is given by
\be
\label{X}
X = X_c + \frac{1}{2} \ddot{X}_c (y^2 +z^2) +O(4).
\ee
Direct substitution of this expression into Eq. (\ref{static}) and
(\ref{static2}) shows it to be a solution, up to the required order,
as long as $y$ and $z$ are always small compared to $\sim \dot{X}^{-1}$.  This
condition should always be true as long as each individual cell is small
compared to the Hubble scale of the lattice, which is the situation
we outlined to begin with.

Now, if $\phi$ is a solution to the Newtonian Laplace equation,
(\ref{statica}), then one may expect the solutions of Eq. (\ref{Xc})
to obey the classification scheme of the usual Newtonian $n$-body
problem given by Saari \cite{saari}.  Roughly speaking, this corresponds to
the following:  If $dX/dt$ is large enough then $X \rightarrow t$ at late-times,
otherwise $X \rightarrow t^{2/3}$ or we have eventual recollapse.
More directly, at some time $t=t_0$ one can place an initial boundary at $x=X(t_0,y,z)$ and
give it some initial velocity $\dot{X}(t_0,y,z)$.
Eqs. (\ref{static}) and (\ref{static2}) then tell us what
$\phi_{,x}\vert_X$ and $\ddot{X}$ are along the boundary at $t=t_0$.
This is enough information to evolve the boundary forward in time,
obtaining $X$ and $\phi_{,x}\vert_X$ at every $t$,  and for every $y$ and
$z$.  It is therefore also sufficient to provide the necessary Neumann boundary
conditions with which we can solve Eq. (\ref{statica}) for
$\phi(x^{\mu})$ at every $t$.  Of course, this can only be done in
ways that satisfy Eq. (\ref{staticb}), and the required properties of
the solutions of Eq. (\ref{statica}).  In particular, the `maximum
principle' of harmonic functions tells us that as long as gravity is
attractive near the central mass then the boundary cannot accelerate.

\section{Fluctuations About Time-dependent Backgrounds}
\label{timesec}

\subsection{Large-scale Evolution Equations}

We can also consider modelling the space-time inside each
cell as fluctuations about time-dependent FRW backgrounds, such that
the line-element can be written as
\be
\label{2}
ds^2=-(1+2 \Phi)d\hat{t}^2+\frac{a^2 (1-2\Psi)}{[1+\frac{k}{4}
  (\hat{x}^2+\hat{y}^2+\hat{z}^2)]^2} (d\hat{x}^2+d\hat{y}^2+d\hat{z}^2), 
\ee
where $a=a(\hat{t})$ and $k=$constant. We are using hatted coordinates
  here, and capital $\Phi$ and $\Psi$, to distinguish these quantities
  from the coordinates and potentials used in the previous section.

Substituting Eq. (\ref{2}) into the Einstein equations $R_{\hat{a}
  \hat{b}}=0$ gives us that $a$ and $\Phi$ must satisfy the following
  equations in the bulk of the cell
\be
\label{FRW}
\frac{a_{,\hat{t}}^2}{a^2} =-\frac{2 \hat{\nabla}^2 \Phi}{3 a^2} -
\frac{k}{a^2} +O(4)
\qquad \mathrm{and} \qquad 
\frac{a_{,\hat{t}\hat{t}}}{a} =\frac{\hat{\nabla}^2 \Phi}{3 a^2} +O(4),
\ee
where $(a \hat{\nabla} \Phi)_{,\hat{t}}=0+O(3)$, where $\Psi=\Phi$ and
where $\hat{\nabla}=\partial_{\hat{\mu}}$ is the spatial derivative operator in the hatted
coordinate system. These are clearly just the Friedmann equations,
with $\hat{\nabla}^2 \Phi/a^2 \propto 1/a^{3}$ acting as a
pressure-less dust term.  If $\hat{\nabla}^2 \Phi=0$, as in the
previous section, then the background space-time is Milne. More
generally $a(\hat{t})$ behaves as in an FRW universe with dust
and spatial curvature.  From Eq. (\ref{FRW}) it can also be seen that we should assign to $k$ an
$O(\epsilon^2)$ of smallness\footnote{We could
  choose to rescale $k$ to $0$ or $\pm 1$, but in this case we would
  also have to rescale all other dimensionful quantities, so the
  overall perturbative expansion would remain unchanged.}.

As before, let us again consider a cell boundary at
$\hat{x}=\hat{X}(\hat{t},\hat{y},\hat{z})$.  The extrinsic curvature
of such a 2+1 dimensional surface is now given, in the coordinates of
Eq. (\ref{2}), by
\bea
\label{Kfrw1}
\hspace{-60pt}
K_{\hat{a} \hat{b}} dx^{\hat{a}} dx^{\hat{b}} 
&=& \left( n_{\hat{t},\hat{t}}-\frac{1}{a^2} n_{\hat{\mu}} \Phi_{,\hat{\mu}} \right) d\hat{t}^2  
+ \left( n_{\hat{\mu},\hat{t}}+n_{\hat{t},\hat{\mu}} -2 \frac{a_{,\hat{t}}}{a} n_{\hat{\mu}} \right) dx^{\hat{\mu}} d\hat{t} 
\\ && \nonumber
+\Bigg[ n_{\hat{\mu},\hat{\nu}} +\left(k x^{\hat{\mu}}+2
  \Psi_{,\hat{\mu}}\right) n_{\hat{\nu}} - \delta_{\hat{\mu}
  \hat{\nu}} \left( \left(\frac{1}{2}k x^{\hat{\sigma}}+
  \Psi_{,\hat{\sigma}}\right) n_{\hat{\sigma}}+ a a_{,\hat{t}} n_{\hat{t}}
  \right) \Bigg] dx^{\hat{\mu}} dx^{\hat{\nu}},
\eea
where quantities with hatted coordinates should be taken to correspond
to those associated with the time-dependent geometry given in
Eq. (\ref{2}).  We can now proceed as in the previous case, making
analogous definitions, to find the extrinsic curvature on a boundary at
$\hat{x}=\hat{X}(\hat{t},x^{\hat{A}})$ to be
\bea
\hspace{-50pt}
K_{\hat{i}\hat{j}} d\xi^{\hat{i}} d\xi^{\hat{j}} &=& - n_{\hat{x}}
\Bigg[ \left(\ddot{\hat{X}}+2\frac{\dot{a}}{a} \dot{X}
  +\frac{1}{a^2}\Phi_{,\hat{x}} - \frac{1}{a^2} \hat{X}_{\vert \hat{A}} \Phi_{,A} \right) d\hat{t}^2
\nonumber \\ && 
\qquad \;\;\;
+ \left( (\hat{X}_{\vert \hat{A}})\dot{\;} + (\dot{\hat{X}})_{\vert \hat{A}} -2
\frac{\dot{a}}{a} \hat{X}_{\vert \hat{A}}\right) dx^{\hat{A}} d\hat{t}
\nonumber \\ &&
\qquad \;\;\;+ \Big( \hat{X}_{\vert \hat{A}\hat{B}} -
\mathcal{J}(\hat{t},\hat{X},x^{\hat{A}}) \left( \delta_{\hat{A}\hat{B}} +\hat{X}_{\vert \hat{A}}
\hat{X}_{\vert \hat{B}}\right) \Big) dx^{\hat{A}} dx^{\hat{B}}
\Bigg],
\label{Kfrw2}
\eea
where we have defined
\be
\mathcal{J}(\hat{t},\hat{X},x^{\hat{A}}) \equiv a \dot{a}
\dot{\hat{X}}- \Psi_{,\hat{x}} + \Psi_{,\hat{A}} \hat{X}_{\vert
  \hat{A}} - \frac{1}{2} k \hat{X} + \frac{1}{2} k x^{\hat{A}}
\hat{X}_{\vert \hat{A}}.
\ee
The junction conditions given in Eq. (\ref{K0}) now tell us that the
boundary must satisfy the following equations, to lowest order in our
expansion:
\bea
\label{c}
\frac{\ddot{\hat{X}}}{\hat{X}} &=& -2
\frac{\dot{a}}{a}\frac{\dot{\hat{X}}}{\hat{X}} 
-\frac{\Phi_{,\hat{x}}}{a^2\hat{X}} + \frac{\hat{X}_{\vert \hat{A}}
  \Phi_{,\hat{A}}}{a^2 \hat{X}}+O(4)
\eea
and
\bea
\label{d}
\hat{X}_{\vert \hat{A}\hat{B}} &=&
\mathcal{J}(\hat{t},\hat{X},x^{\hat{A}}) \left( \delta_{\hat{A}\hat{B}} +\hat{X}_{\vert \hat{A}}
\hat{X}_{\vert \hat{B}}\right)+O(4)
\eea
with $(\hat{X}_{\vert \hat{A}})\dot{\;}=\Big( \dot{\hat{X}} \Big)_{\vert \hat{A}}
=2 (\dot{a}/a)  \hat{X}_{\vert \hat{A}} +O(3)$.  These equations must be
satisfied, together with the bulk field equations, (\ref{FRW}), in
order to have a viable global solution to Einstein's equations.

\subsection{Cosmological Solutions}

Eqs. (\ref{FRW}), (\ref{c}) and (\ref{d}) admit as a solution a
lattice cell with boundaries that are static in the
coordinates of Eq. (\ref{2}).  In this case, $\hat{X}$ must satisfy
\be
\label{Xfrw}
\hat{X}_{\vert \hat{A} \hat{B}} = -\frac{k}{2} \left( \hat{X} -
\hat{X}_{\vert \hat{C}} x^{\hat{C}} \right) \left(
\delta_{\hat{A}\hat{B}}+X_{\vert \hat{A}} X_{\vert \hat{B}} \right) +O(4),
\ee
for all $\hat{t}$, which has the solution
\be
\label{Xfrw2}
\hat{X} = \hat{X}_0 \left( 1-\frac{k}{4} \left(
\hat{y}^2+\hat{z}^2\right) \right) +O(4),
\ee
where $\hat{X}_0$ is a constant.  As $\hat{X}$ is not a function of $t$ here, the global expansion of the
space-time is fully specified by $a(\hat{t})$ alone.  Eq. (\ref{c})
then tells us that for Eq. (\ref{Xfrw}) to hold we must require
\be
\label{con}
\hat{n} \cdot \hat{\nabla} \Phi \vert_{\hat{X}} =0+O(4)
\ee
on the boundary.  Now, it also clear from Eq. (\ref{FRW}) that the
potential $\Phi$ must satisfy
\be
\label{Phi}
a \hat{\nabla}^2 \Phi = c_1+O(4),
\ee
where $c_1$ is a constant in both $\hat{t}$ and $x^{\hat{\mu}}$, and
is $O(2)$ in smallness.  This equation has the solution
\be
\label{psi}
\Phi = \frac{\Phi_N}{a} + \frac{c_1}{6 a}
(\hat{x}^2+\hat{y}^2+\hat{z}^2) +O(4),
\ee
where $\Phi_N=\Phi_N(x^{\hat{\mu}})$ is the Newtonian potential that
satisfies $\hat{\nabla}^2 \Phi_N=0$.  The extra non-Newtonian term
must occur for all time-dependent backgrounds other than the Milne
universe, and ``appears due to the fact that in the present case we
have no embedding in the Euclidean space'' \cite{es}.  We will call
this term the vacuum potential.

The solutions (\ref{Xfrw}) and (\ref{psi}) above can be seen to
satisfy the condition (\ref{con}) if, and only if,
\be
c_1 = - \frac{3}{\hat{X}} \Phi_{N,x} \vert_{\hat{X}}.
\ee
One can then see that if the Newtonian
potential is attractive, so that $\Phi_{N,\hat{x}} >0$, then
boundaries stay at fixed $\hat{X}$ if, and only if, $c_1<0$.
It can also be seen from Eq. (\ref{Xfrw2}) that for $k=0$ a boundary
at static $\hat{X}$ must be in a plane of $\hat{x}=$constant, while for
$k<0$ or $k>0$ the boundary must be either convex or concave, respectively,
when viewed from inside the cell in the coordinates of Eq. (\ref{2}).
It is also clear from Eq. (\ref{FRW}) that the functional form of $a(\hat{t})$ in
each of these cases must be the same as in an FRW universe with the
same $k$.

Let us also note that solutions with boundaries at static $\hat{X}$ require the
forces from the Newtonian and vacuum potentials in 
Eq. (\ref{psi}) to be in unstable equilibrium.  This is due to the
force from the Newtonian potential being attractive and growing as $\hat{X}$ becomes
smaller, while the force from the vacuum potential pushes the boundary
to greater $\hat{X}$ (as long as $c_1<0$), and grows with $\hat{X}$.
So, if the boundary should be perturbed to slightly large $\hat{X}$, then
the vacuum potential should come to dominate, and cause
$d^2 \hat{X}/d\hat{t}^2>0$. This does not, however, correspond to acceleration in the more
cosmologically relevant proper distance $R=a \hat{X}$, which from Eqs.
(\ref{FRW}), (\ref{c}) and (\ref{psi}) can be seen to be given by
$d^2 R/d\hat{t}^2 =-a^{-2} \Phi_{N,\hat{x}}\vert_{\hat{X}}$.  As in the case of static
backgrounds, the expansion of the global lattice is therefore always
decelerating (in terms of the proper distance, $R$), as long as the
force from the Newtonian potential is attractive.

\section{Relating Time-dependent \& Time-independent Descriptions}
\label{7}

One could now ask if it is possible to remove the vacuum potentials by
transforming the $\hat{X}=\hat{X}(\hat{y},\hat{z})$ solutions above to
different homogeneous and isotropic backgrounds.  We find that this is
indeed possible under the coordinate redefinitions
\bea
\hat{t} &=& t -\frac{a_{,t}(t)}{a(t)} \frac{(x^2 +y^2
  +z^2)}{2} +O(3)\\
\hat{x} &=& \frac{x}{a(t)} \left[1- \left(\frac{c_1}{6 a^3(t)}
  +\frac{k}{4 a^2(t)} \right) (x^2+y^2+z^2)  \right] +O(4) \\
\hat{y} &=& \frac{y}{a(t)} \left[1- \left(\frac{c_1}{6 a^3(t)}
  +\frac{k}{4 a^2(t)} \right) (x^2+y^2+z^2) \right] +O(4) \\
\hat{z} &=& \frac{z}{a(t)} \left[1- \left(\frac{c_1}{6 a^3(t)}
  +\frac{k}{4 a^2(t)} \right) (x^2+y^2+z^2)  \right] +O(4).
\eea
The argument of $a$ is made explicit here, as it is now the case that
$a(\hat{t}) \neq a(t)$.  Instead we have to Taylor expand to find
\be
a(\hat{t}) = a(t)\left( 1- \frac{a_{,t}^2(t)}{a(t)}
\frac{(x^2+y^2+z^2)}{2} \right) +O(4).
\ee
The line-element that results is then the static one specified in
Eq. (\ref{1}), with 
\be
\phi= \frac{\Phi_N}{a(\hat{t})} +O(4).
\ee
We can then see that a boundary at $\hat{X}=\hat{X}_0 (1-
(\hat{y}^2+\hat{z}^2) k/4) +O(4)$, in the coordinates of Eq. (\ref{2}), is
equivalent to the solution given in Eq. (\ref{X}) with 
\be
X_c(t) = a(t) \hat{X}_0 \left( 1+ \frac{k \hat{X}_0^2}{4} \right) +O(4).
\ee
The constant $k$ then determines the asymptotic form of $X_c(t)$ in
the usual way that is familiar from the solutions to the Friedmann equation.

We have now shown that one can describe the space-time geometry
inside each of our cells as either linearly perturbed FRW, as in Eq. (\ref{2}), or as
linearly perturbed Minkowski space, as in Eq. (\ref{1}), and that these two descriptions
are equivalent to each other up to a coordinate
transformation.  One could argue, however, that the more natural
description is in terms of the static coordinate system of Eq. (\ref{1}).  In
these coordinates the perturbations can be consistently described
as being solely due to Newtonian potentials.  This is not true in the
FRW coordinates, where an additional potential of the form $\Phi \sim r^2$
is also required within each cell (unless one wants to use an open, empty
Milne universe as the background cosmology).  
If one were to choose to use the FRW description, it would therefore seem
necessary to understand the effects that these potentials have on
observable quantities, which are discussed further in Section \ref{obs}.
Using the Newtonian (static) description, for calculations of
luminosity distances etc., however,
requires taking into account boundary condition between
the different regions in which Newtonian descriptions are internally
applicable.  This will also be discussed in Section \ref{obs}.

\section{Comparison with Perfect Fluid FRW Cosmology}
\label{8}

As discussed in Section \ref{2b}, it has been suggested that the
formation of structure in the Universe could lead to modifications of
the expected FRW cosmological expansion, and in particular the scale of expansion.
We will now use the model described in the previous section to address this issue.
As already shown, the functional form of the constraint and evolution
equations are the same as the Friedmann equations of FRW.  This does
not, however, guarantee that the solutions of these equations are
identical to FRW solutions.  In particular, we need to verify that the
energy density corresponding to a given expansion rate is the same as
expected from FRW, and that in spatially curved solutions the scale of
expansion is also as expected. 

\subsection{Spatially Flat Cosmologies}

\begin{figure}[tbp]
\centerline{\epsfig{file=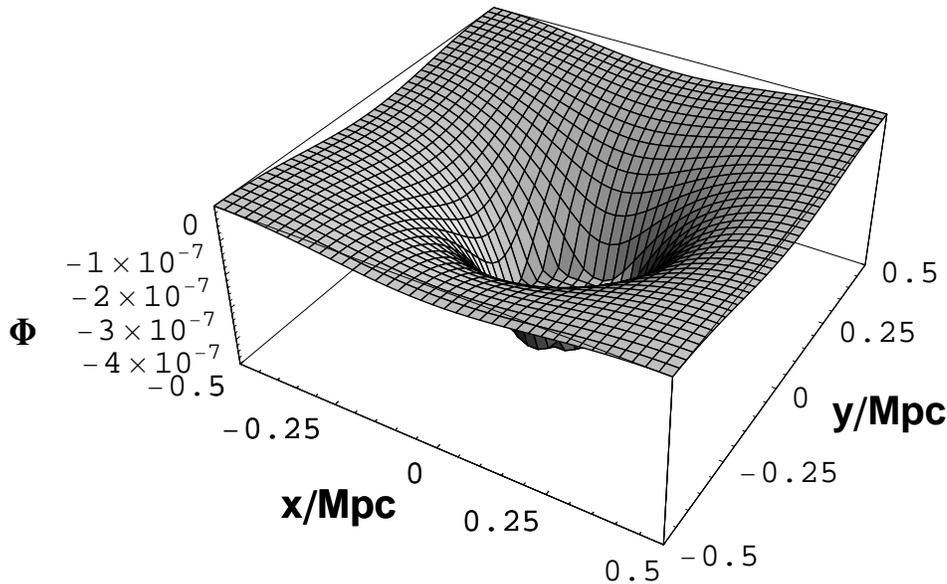,height=7.5cm}}
\caption{\textit{The potential $\Phi$ satisfying the field equation
    (\ref{phifield}), and the boundary conditions (\ref{phiboundary})
    with a central mass of approximately Milky Way size, in a $1$Mpc
    sized cubic cell, and in the plane $z=0$.}}
\label{phifig}
\end{figure}

First of all let us consider a spatially flat model.  In this case $a(t)$
is scale-invariant, and so we only need to check that the value of
the expansion rate for a given energy density is the same as prescibed
by the usual Friedmann equation.  To do this, we will
find numerical solutions for the potentials $\phi$ and $\Phi$ that satisfy
\be
\label{phifield}
 \nabla^2 \phi= \hat{\nabla}^2 \Phi - \frac{c_1}{a}  = O(4),
\ee
with boundary conditions given by 
\be
\label{phiboundary}
(n \cdot \nabla \phi) \vert_X + X_{\vert AA} =
\hat{n} \cdot \hat{\nabla} \Phi \vert_{\hat{X}} = O(4),
\ee
and with a singularity at the centre of the cell.  This solution is shown
in Figure \ref{phifig}, for a central mass approximately as
massive as the Milky Way 
, in a cubic cell of width $1$Mpc.  We can then verify that, for
fields satisfying Eqs. (\ref{phifield}) and (\ref{phiboundary}), we have
\bea
\label{phinum1}
(n \cdot \nabla \phi) \vert_X \simeq -\frac{4 \pi}{3} \rho X,\\
\label{phinum2}
\hat{\nabla}^2 \Phi \simeq -4 \pi a^2 \rho,
\eea
where $\rho=m/V$, $m$ is the central mass, and $V$ is the spatial
volume of the cell (in either coordinate system, to the required
accuracy).  The $\simeq$ sign here means equal up to terms of $O(4)$.  Substituting
Eq. (\ref{phinum1}) into (\ref{static}), or Eq. (\ref{phinum2}) into
(\ref{FRW}), we then recover the usual Friedmann equation, up to corrective
terms of $O(4)$.  The expansion rate for a given density is
therefore the same as in a perfect FRW universe, to the required
order.
This result can also be obtained from Gauss's theorem by noting that
from Eqs. (\ref{con}) and (\ref{Phi}) we have
\bea
\frac{1}{a^2} \int_V \hat{\nabla}^2 \Phi dV &= \frac{1}{a^2} \oint
\hat{n} \cdot (\hat{\nabla} \Phi ) dS = O(4),\\
&= \frac{c_1 V}{a^3} +4 \pi m +O(4)
\eea
where integrals are performed over the volume of a cell, $V$, enclosed by the surface,
$S$, and $m$ is the mass of a central singularity.  It is then clear
that $c_1 = -4 \pi \rho a^3 +O(4)$, which on substitution into
Eq. (\ref{FRW}) again gives the usual Friedmann equation, up to terms of $O(4)$.

Finally, we point out that the $\rho$ occurring in Eqs. (\ref{phinum1}) and (\ref{phinum2})
does not correspond to a local energy density, and hence that these
equations should not be considered as actual field equations
themselves (as they are in studies of perturbed FRW solutions, with
perfect fluids), but rather as a derived consequence of
the field equations.  What has then be shown above is that the
large-scale evolution of a spatially flat universe can be recovered, up to some
prescribed level of accuracy, even when the matter content is arbitrarily
inhomogeneous (at least, when the matter is arranged in a regular
way).  This result relies only on the validity of local perturbative
expansions about Minkowski space in the vicinity of isolated masses,
which is not often considered an ambiguous procedure.  This is not the
same thing as taking for granted the validity of the usual perturbed
FRW approach and showing that regions of space-time can then be
described locally as perturbed Minkowski space.  In that case the
large-scale expansion is given by the assumed global background,
rather than as a consequence of any boundary conditions.


\subsection{Non-flat Cosmologies}

Now let us consider models with non-zero spatial curvature.  In these cases the
corrections to Eqs. (\ref{phifield}) and (\ref{phiboundary}) due to
the curvature are of $O(4)$ only, and hence do not affect our
numerical calculation of $\phi$.  Furthermore, the RHS of
Eqs. (\ref{phinum1}) and (\ref{phinum2}) are also only corrected by
$O(4)$ terms.  It is therefore the case the energy density in the
spatially curved models is also the same as in the usual FRW Friedmann
equations, up to the required order of accuracy.  It now remains to confirm this
is also true for the spatial curvature terms in Eqs. (\ref{static})
and (\ref{FRW}).

In order to compare the scale of expansion in the models we have been
considering to the usual FRW perfect fluid solutions we need to know
if the $k$ appearing in Eq. (\ref{FRW}) is the same as $K$, which
determines the global spatial curvature of the homogeneous FRW solutions.  Global
curvature in the lattice models has {\it not} yet been shown to be
equal to $k$, which so far only describes spatial curvature inside of
each cell in the coordinates used in Eq. (\ref{2}), and the position
of the boundaries via Eq. (\ref{Xfrw2}).  

Global curvature in the lattice models should be inferred using the
angle at which the different faces of a single cell meet.  To see
this, consider a lattice made from cells that are a single repeated regular polytope.
the resulting structure is then known as a polychoron, and there are 6
different convex polychora with which we could model a lattice with
positive spatial curvature \cite{coxeter}. Now, take as an example the
largest of these configuration, which contains 600 simplexes, and is known as a
hexacosichoron.  It has 5 simplexes meeting around each edge of every
face of every cell, and so in order to create such a structure we would need the angle
with which the cell boundaries described by Eq. (\ref{Xfrw2}) meet to
be $360^{\circ}/5=72^{\circ}$, in a hyper-surface orthogonal to the
world-line of a time-like observer on the edge of the cell face.
This angle can be seen to depend on the value of
$k$ in Eq. (\ref{Xfrw2}), and once it has been achieved the global
curvature of the lattice is set by the curvature $K$ of the hyper-sphere
with an image of the same polychoron on its surface volume.

We can now note that the boundary positions of each cell, when
considered as an image on a global hyper-sphere, should be described
by geodesics of that hyper-sphere, due to the symmetry about each
boundary.  Ensuring that Eq. (\ref{Xfrw2})
describes a geodesic in such a space therefore allows us to compare
$k$, from the boundary position equation, with $K$ from the geodesic
equations on the hyper-sphere.  To do this, we differentiate Eq. (\ref{Xfrw2}) twice with
respect to some affine parameter, $\lambda$, along a curve in the boundary, giving
\be
\label{ddX}
\ddot{X} = -\frac{k}{2} X_0 \left( \dot{y}^2 +\dot{z}^2+y \ddot{y}+z
\ddot{z} \right) +O(4),
\ee
where over-dots here denote differentiation with respect to $\lambda$.
Using the metric of a hyper-sphere with spatial curvature $K$,
\be
\label{hsmet}
ds^2=\frac{dx^2+dy^2+dz^2}{(1+\frac{K}{4}(x^2+y^2+z^2))^2},
\ee
we then find that in order for $\ddot{X}$, $\ddot{y}$ and $\ddot{z}$
to describe a geodesic we must have
\bea
\ddot{X} &=& -\frac{K X}{2} \left( \dot{y}^2 + \dot{z}^2 \right) +O(4),\\
\ddot{y} &=& \frac{K y}{2} \left( \dot{y}^2 - \dot{z}^2 \right) +K z
\dot{y} \dot{z} +O(4)\\
\ddot{z} &=& \frac{K z}{2} \left( \dot{z}^2 - \dot{y}^2 \right) +K y
\dot{y} \dot{z} +O(4),
\eea
where we have used the result $X_{\vert A} \sim O(1)$ to assign $\dot{X}$ an
$O(1)$ of smallness, compared to $\dot{y}$ and $\dot{z}$. Substituting these expressions into Eq. (\ref{ddX}) then gives
\be
\label{kK}
k=K+O(4).
\ee
We have therefore shown that the boundaries described by
Eq. (\ref{Xfrw2}) are, in fact, geodesics of the hyper-sphere given in
Eq. (\ref{hsmet}) when Eq. (\ref{kK}) is satisfied. The global
curvature of lattice must therefore be given by $k$, with only
corrections up to $O(4)$ allowed.

With the energy density and spatial curvature terms on the RHS of
Eq. (\ref{FRW}) being equal to their values in the corresponding FRW
solutions, the scale of expansion of the lattice model must therefore also be
equal to that of the FRW solutions.  We
have now shown that the large-scale dynamics of these models are indistinguishable
from those of perfect fluid FRW solutions with the same global energy density
and spatial curvature, up to the required accuracy.  As we have also shown
that the geometry inside each cell can be described as perturbed FRW
geometry, it then follows that the background FRW solution of each cell can also be
taken to be the global solution.  We have therefore shown that a global perturbed FRW
space-time geometry is sufficient to describe the situation of n
regularly spaced discrete masses, with no corrections beyond $O(4)$
required (unless one wants to describe the region close to a compact
object, or the early universe).

\section{Observables in a Lattice}
\label{obs}

We have so far considered the space-time geometry of a universe
composed of a large number of discrete objects, nearby to each of
which post-Newtonian gravity is a good approximation.  The large-scale
evolution of the global space-time has then been deduced through
the applications of Israel junction conditions between the different
local patches, and it has been found that the usual linearly perturbed FRW
cosmology is still a good approximation to the space-time geometry (as
long as one does not approach a singularity) even though the density
fluctuations are arbitrarily large.  These results are promising
evidence for the applicability of perturbed FRW cosmology to at least
some situations in which the density contrast is large. It does not,
however, guarantee that cosmological observations in these space-times
will be similar to those made in a perfect FRW space-time.  We will
consider this problem in the present section.

There are, of course, a great number of studies on observable
quantities such as redshift and luminosity distance in inhomogeneous
cosmological models.  This is particularly true in the case of
perturbed FRW cosmology, where the relevant formalism was first given by Kristian and Sachs
\cite{ks}.  With regards to other approaches, the `Swiss cheese'
models, in which Einstein-Strauss vacuoles \cite{es} or spherical Lema\^{i}tre-Tolman-Bondi
patches \cite{ltb1}-\cite{ltb3}, are embedded in a perfect FRW
background have also been well studied \cite{kant}-\cite{cz}.  Such
studies allow for the contribution of non-linear and non-perturbative
effects, but only for spherically symmetric inhomogeneities with
perfect FRW boundary conditions.  Observables in non-FRW models with regularly spaced
discrete masses were studied in \cite{cf1} and \cite{cf2}, using the
Lindquist-Wheeler model.

Here one could proceed with calculating redshifts and luminosity
distances in at least two different ways:  (i) Within the context of the
perturbed FRW geometry given by Eq. (\ref{2}), or (ii) in terms of the
static geometry given by Eq. (\ref{1}).  In either case the results
should be the same, and the first
step is to calculate the photon trajectories within each cell.  These
should satisfy the geodesic and null constraint equations
\be
\nonumber
k^a k^b_{;a} = 0 \qquad \qquad {\rm and} \qquad \qquad k^a k_a = 0,
\ee
where $k^a$ is the 4-vector tangent to the null geodesics.  In either
of the two cases mentioned above one must then deal with the boundary conditions
between cells, in order to propagate photons over cosmologically
relevant distance scales.  This should proceed as follows:  One should
define a congruence of time-like geodesics that are comoving with the
boundary of the first cell, which can be labelled $u^{a_1}$.  One can then decompose $k^{a_1}$
into components tangential and orthogonal to $u^{a_1}$ as
\be
\label{kdec}
k^{a_1} = (-u^{b_1} k_{b_1}) (u^{a_1}+n^{a_1}),
\ee
where $n^{a_1} u_{a_1} =0$, and $n^{a_1} n_{a_1}=1$.  The frequency and direction of a
photon that passes by an observer on the boundary are then given by
$-u^{a_1} k_{a_1}$ and $n^{a_1}$, respectively.  Here we have used
subscript $1$ on space-time indices to denote that these quantities
are being evaluated in the coordinates used in the first cell.

Now consider an observer on the boundary of the second cell following
a curve in the time-like congruence $u^{a_2}$, into which the photon is
propagating.  A similar decomposition to Eq. (\ref{kdec}) can be
performed, and the frequency and direction of the photon measured by
this observer are given by $-u^{a_2} k_{a_2}$ and $n^{a_2}$.  As the
observer on the boundary of the first cell is to be identified with the
observer on the boundary of the second cell, when the junction
conditions are applied, the frequency and direction of photons measured by the two
observers should be the same.  The quantities $-u^{a_2} k_{a_2}$ and
$n^{a_2}$ are therefore given by
\bea
-u^{a_2} k_{a_2} = -u^{a_1} k_{a_1}\\
n^{a_2} = \frac{\partial x^{a_2}}{\partial x^{a_1}} n^{a_1},
\eea
where $({\partial x^{a_2}}/{\partial x^{a_1}})$ denotes the relevant
transformation between the two coordinate systems.  These four
equations provide enough information to calculate the four components
of $k^{a_2}$, when $k^{a_1}$ is known.  We therefore have the initial
conditions we need to propagate the congruence of null geodesics
through the second cell, where the same procedure as just described
can be applied again and again to propagate through large numbers of
cells.

With a knowledge of the 4-vector $k^a$ we can now calculate
cosmological redshifts and luminosity distances in the model under
consideration.  The first of these quantities is given by the
expression
\be
\label{rs}
1+z = \frac{(- u^a k_a)\vert_{e}}{(-u^b k_b)\vert_{o}},
\ee
where, as already explained, $- u^a k_a$ corresponds to the frequency
of a photon measured by an observer following a curve in $u^a$, and we
have used subscripts $o$ and $e$ to denote the points on the null curve
where observation and emission occur.  There is some ambiguity here in
exactly how one chooses the congruences $u^a$, as the space-time
is a vacuum outside of the central object of each cell, and so no
preferred set of curves given by a background fluid can be assumed.
As long as one is consistent in how such congruences are chosen from
cell to cell, however, the differences should be small between the
different possible choices, as long as the relative velocity between
two observers following the curves in the two congruences is also small.

Finally, in order to obtain luminosity distances, one needs to
integrate the Sachs optical equations along the null curves that were
found above.  These are:
\bea
\label{ex}
\frac{d \theta}{d \lambda} + \theta^2-\omega^2+\sigma^*\sigma =
-\frac{1}{2} R_{ab} k^a k^b\\
\frac{d\omega}{d\lambda} +2 \omega \theta = 0\\
\label{ro}
\frac{d\sigma}{d\lambda} + 2 \sigma \theta = C_{abcd} (t^*)^a k^b
(t^*)^c k^d,
\label{sh}
\eea
where $\theta$, $\omega$ and $\sigma$ are the expansion, rotation and
complex shear scalars, respectively, $\lambda$ is an affine parameter
along the curve, and $t^a$ is a complex vector
field obeying $t^a k_a=0$, $t^a t_a=0$ and $t^a (t^*)_a=1$.  Here, $R_{ab}$
is the Ricci tensor and $C_{abcd}$ is the Weyl tensor.  In propagating
photon trajectories between cells one should transform $t^a$ in a similar way to $k^a$.
As always, once the $\theta$, $\omega$ and $\sigma$ are known along the null
trajectories, then the angular diameter distances can be calculated by
$
r_A \propto \rm{exp} \left\{ \int^o_e \theta d \lambda \right\},
$
and the luminosity distances by
$
r_L = (1+z)^2 r_A.
$

We will now make some comments about the procedure outlined above,
which will not be performed explicitly here.  Firstly, we note that
in perfect fluid FRW cosmology the RHS of
Eq. (\ref{ex}) is non-zero and the RHS of Eq. (\ref{sh}) is zero,
while in the present situation the exact opposite is true:  The RHS of
Eq. (\ref{ex}) is zero and the RHS of Eq. (\ref{sh}) is non-zero.
This appears to have been first noticed by Bertotti in \cite{bert}.
Now, while the Ricci curvature source term is switched off in the
present case, the $\sigma^*\sigma$ term in Eq. (\ref{ex}) becomes
non-zero, due to the Weyl term on the RHS of Eq. (\ref{sh}) being
non-zero.  It has been argued by Weinberg that the effect of the
non-zero shear in Eq. (\ref{ex}) should entirely replace the missing
Ricci curvature term, so that the luminosity distance is, in fact,
unaffected by the matter being clumped \cite{wein}.  This argument is
essentially based on conservation of photon number, but has since been
shown to be questionable \cite{ebd}.  In particular, Weinberg's
argument neglects the occurrence of caustics in the congruence of null
geodesics along which we observe.  Caustics should be expected to
occur when shear is allowed to be non-zero, as it necessarily has to be here.
Furthermore, Weinberg's argument relies on spatial areas taking the same value
in clumpy cosmological models as they do in FRW ones, and on being
able to make large numbers of observations over the entire sky.  
With finite observations made at relatively low redshifts ($z \lesssim 1$),
shear has in fact been shown, for at least some inhomogeneous
cosmological models, to remain relatively low along typical 
geodesics \cite{cf1,mar2}.  We expect this to be true in the present
situation as well, so that $\theta \simeq 1/\lambda$ when $z \lesssim
1$.  The neglect of shear in this way is sometimes known as the
Dyer-Roeder approximation \cite{dr}, and while it is expected to be a good
approximation for most situations at low redshifts, it may not be so
at high redshifts.  The effect that shear can have on the CMB sky has been
discussed in \cite{ebd}, where the consequences of caustics in
particular are considered.

It now remains to consider redshifts along null geodesics.  These
are usually taken as being prescribed by the ratio of scale factors at
different points on a curve, as in FRW cosmology, although this does
not necessarily have to be the case in inhomogeneous models, and in
general one should calculate it using Eq. (\ref{rs}).  In the
Lindquist-Wheeler model it was recently found that a non-negligible
deviation from the FRW value of redshift can occur \cite{cf1}.  For
the model being considered here, we do not expect a repeat of this
result.  The reason for this is that we can choose the time-like
congruence $u^a$ to be that of an observer at fixed spatial position,
in the coordinates of Eq. (\ref{2}).  Observers following such a
congruence will not be geodesic, but this does not matter, and it
should only be a small correction if one wishes to consider geodesic
observers, as peculiar velocities in the coordinates
of Eq. (\ref{2}) are all $O(1)$ small.  Now, if the 4-vector tangent
to the null geodesics, $k^a$, were the same as in FRW then we would
expect exactly the same redshift along each curve.  Here we should
expect small perturbations to this field, so the actual redshift
should also be corrected.  Assuming the photon
trajectories do not pass too close to the Schwarzschild radius of any
central masses, however, these corrections are expected to be small, as is the
case in Swiss cheese models \cite{kant}.  We therefore do not expect
large deviations from the expected FRW results on redshift, as were
found in \cite{cf1}, although one would need to perform the explicit
numerical integration described above to be sure.

The picture we have outlined here is quite similar to that used by Holz and
Wald in their ``new method for determining cumulative gravitational
lensing effects'' \cite{hw}.  In this work the authors use Newtonian
potentials to calculate the shear and expansion of a bundle
of null geodesics as it passes through a region, updating the redshift
using FRW relations as they leave one region and enter the next.  This
is just the situation here:  Inside each cell the space-time can be well
described by the perturbed Minkowski space of Eq. (\ref{1}), and by
propagating geodesics between cells one should pick up a similar
redshift to that expected from FRW cosmology (up to possible
effects caused by perturbations to $k^a$, as just discussed).

\section{Discussion}
\label{disc}

In summary, we have considered n-body cosmological models that do not require any
averaging procedures.  These models have been constructed using a
lattice made from a large number of symmetric cells, each containing identical
central masses.  By applying appropriate junction conditions between cells
we then find a set of evolution equations that specify the motion of
the cell boundaries, and hence the global evolution of the space-time.
In all of the cases we have considered we find that the cell boundaries
must be in free fall, and decelerating in their expansion away from
the cell centres.  We find that the space-time geometry inside each
cell can be described as fluctuations around either static or
time-dependent FRW backgrounds.  In order to apply the FRW description
one must include potentials of the form
$\Phi \sim r^2$, whose gradients must be of the same order of magnitude as the
Newtonian potentials that are also present.  Treating the
space-time inside each cell as fluctuations around Minkowski space,
however, requires Newtonian potentials only.  One may then consider
perturbed Minkowski space to be a more natural local description of the space-time.

We have compared the resulting large-scale dynamics of the
cosmological model under consideration with those of a perfect fluid FRW
cosmology.  We find that for spatially flat universes the expansion
rate for a given energy density has just the expected FRW value.
Furthermore, we have compared the scale of expansion of non-flat
models with spatially curved perfect fluid FRW universes and found that they are
also the same, up to the required accuracy.  We emphasize that this
result does {\it not} follow directly from the fact that the
space-time geometry inside each cell can be written as perturbed FRW:
It could, in principle, have been the case that the scale
of curvature of the global lattice was different from the
scale inferred from the relevant FRW background used inside each cell
(in fact, from a number of the previous studies discussed in Section
\ref{2b}, it would seem that one may have expected such a result).
Instead, what we find here is that discretisation of the matter
content of the Universe does {\it not} have to affect the global background
rate of expansion.

These results are a consequence of explicitly solving boundary
conditions between different regions that can, individually, be
described as perturbations around highly symmetric backgrounds.  That
such a description is possible within each cell does not, however,
automatically mean that one can treat the entire Universe as
fluctuations around a single universal FRW background.  Such a result
has to be shown to be true by reconstructing the global geometry from the local
geometry that is appropriate within each of these regions.  This is what we have
done here, finding that for the simple case of regularly spaced masses
a single global FRW background with small perturbations around it is a perfectly
adequately description of the entire space-time.  That a
global FRW background is a valid description in the present case, however, does
not automatically mean that this will also be true for irregular arrangements of
massive objects.  Such a result remains to be shown.

As already discussed, the model we have considered here can be interpreted solely in terms
of perturbations about Minkowski spaces (albeit a different Minkowski
space within each cell).  That this is possible gives one greater
reason to expect our perturbative expansion to be valid in the regime
of non-linear density contrasts than is the case in globally FRW perfect
fluid cosmology.  As discussed in \cite{ras}, there are terms at
higher orders in the latter case that one may expect to blow up when
$\delta \rho/\rho$ becomes large.  With the former case of Newtonian and post-Newtonian
fluctuations about Minkowski space, however, the situation is
different.  The terms that could potentially blow up are absent, and
hence we have greater confidence in the applicability of the
post-Newtonian description of these systems when density contrasts are
large.

To determine whether or not the results we have found here carry over to more
realistic models of the Universe, where masses are irregularly spaced,
and dispersed matter is also present, will require further study, and more
refined models.  However, we do expect our results to be of use for
constraining the possible effects that structure formation could have on the
background expansion of the Universe, as well as for testing the viability of mechanisms that
have been constructed in order to correct for back-reaction and
averaging:  If corrections are predicted in situations where it is
known that none occur, then the frame-work within which they have been
identified should be questioned.

\newpage

\vspace{10pt}

\begin{flushleft}
{\bf Acknowledgements}
\end{flushleft}

I am grateful for valuable discussions with Pedro Ferreira, Chris Clarkson,
George Ellis and Thomas Sotiriou.  I also wish to acknowledge the
support of Jesus College Oxford, the BIPAC, and CERN.




\section*{References}

\begin{thebibliography}{99}


\bibitem{ellis} Ellis G F R 1984 in \textit{General Relativity and
  Gravitation}, Ed. Bertotti B \textit{et al.} Reidel, Amsterdam

\bibitem{buchert} Buchert T 2000 \textit{Gen. Rel. Grav.} \textbf{32}
  105

\bibitem{ras} R\"{a}s\"{a}nen S 2010 {\it Phys. Rev. D} \textbf{81} 103512

\bibitem{mcv} McVittie G C 1931 {\it Mon. Not. Roy. Astron. Soc.}
  \textbf{91} 274

\bibitem{es} Einstein A and Strauss E G 1945 \textit{Rev. Mod. Phys.}
  \textbf{17} 120 

\bibitem{LW} Lindquist R W and Wheeler J A 1957 \textit{Rev. Mod. Phys.} \textbf{29}
432

\bibitem{cf1} Clifton T and Ferreira P G 2009 \textit{Phys. Rev. D}
  \textbf{80} 103503

\bibitem{cf2} Clifton T and Ferreira P G 2009 \textit{JCAP}
  \textbf{10} 26 

\bibitem{larena} Uzan J-P, Ellis G F R and Larena J 2010
  \textit{arXiv}:1005.1809[gr-qc]

\bibitem{zal} Zalaletdinov R M 1997 {\it Bull. Astron. Soc. India}
  \textbf{25} 401

\bibitem{cpz} Coley A A, Pelavas N and Zalaletdinov R M 2005 {\it
  Phys. Rev. Lett.} \textbf{95}, 151102

\bibitem{buch} Buchert T 2008 {\it Gen. Rel. Grav.} \textbf{40} 467

\bibitem{lar1} Buchert T, Larena J and Alimi J-M 2006 {\it
  Class. Quant. Grav.} \textbf{23} 6379

\bibitem{israel} Israel W 1966 \textit{Nuovo Cim. B} \textbf{44} 1

\bibitem{israel2} Israel W 1967 \textit{Nuovo Cim. B} \textbf{48} 463

\bibitem{tegp} Will C M 1993 \textit{Theory and Experiment in
  Gravitational Physics, revised edition}, CUP

\bibitem{saari} Saari D G 1971 \textit{Trans. Am. Math. Soc.}
  \textbf{156} 219

\bibitem{coxeter} Coxeter H S M 1973 {\it Regular Polytopes,
  3rd. ed.}, Dover Publications 

\bibitem{ks} Kristian J and Sachs R K 1966 {\it Astrophys. J.},
  \textbf{143} 379

\bibitem{ltb1} Lema\^{i}tre G 1933 {\it Ann. Soc. Sci. Brussels A}
  \textbf{53} 51

\bibitem{ltb2} Tolman R C 1934 {\it Proc. Nat. Acad. Sci. USA}
  \textbf{20} 169

\bibitem{ltb3} Bondi H 1947 {\it Mon. Not. Roy. Astron. Soc.}
  \textbf{107} 410

\bibitem{kant} Kantowski R 1969 {\it Astrophys. J.} \textbf{155} 89

\bibitem{mar} Marra V, Kolb E W, Matarrese S and Riotto A 2007 {\it Phys. Rev. D}
  \textbf{76} 123004

\bibitem{bis1} Biswas T, Mansouri R and Notari A 2007 {\it JCAP}
  \textbf{0712} 017

\bibitem{bis2} Biswas T and Notari A 2008 {\it JCAP} \textbf{0806} 021

\bibitem{bro1} Brouzakis N, Tetradis N and Tzavara 2007 {\it JCAP}
  \textbf{0702} 013

\bibitem{bro2} Brouzakis N, Tetradis N and Tzavara 2008 {\it JCAP}
  \textbf{0804} 008

\bibitem{cz} Clifton T and Zuntz J 2009 {\it
  Mon. Not. Roy. Astron. Soc.} \textbf{400} 2185

\bibitem{bert} Bertotti B 1966 {\it Proc. Roy. Soc. Lond. A}
  \textbf{294}, 195

\bibitem{eth} Etherington I M H 1933 {\it Phil. Mag. ser. 7}
  \textbf{15}, 761

\bibitem{wein} Weinberg S 1976 {\it Astrophys. J.} \textbf{168} 57

\bibitem{ebd} Ellis G F R, Bassett B A C C and Dunsby P K S 1998 {\it
  Class. Quant. Grav.} \textbf{15} 2345

\bibitem{mar2}  Kainulainen K and Marra V 2009 {\it Phys. Rev. D}
  \textbf{80} 123020

\bibitem{dr} Dyer CC and Roeder R C 1973 {\it Astrophys. J. Lett.}
  \textbf{180} L31

\bibitem{hw} Holz D E and Wald R M 1998 {\it Phys. Rev. D} \textbf{58}
  063501






\end{thebibliography}
\end{document}